\documentclass{PoS}

\title{The dynamical gluon mass in the massless bound-state formalism}

\ShortTitle{The dynamical gluon mass in the massless bound-state formalism}

\author{\speaker{David Ibanez}\\
         European Centre for Theoretical Studies in Nuclear Physics
         and Related Areas (ECT*) and Fondazione Bruno Kessler,\\
         Villa Tambosi, Strada delle Tabarelle 286, I-38123 Villazzano (TN) Italy \\
         
        E-mail: \email{ibanez@ectstar.eu}}

\abstract{We describe the phenomenon of dynamical gluon mass generation within the massless bound-state formalism, which constitutes the general framework for the systematic implementation of the Schwinger mechanism in non-Abelian gauge theories. The main ingredient of this formalism is the dynamical formation of bound states with vanishing mass, which gives rise to effective vertices containing massless poles; these vertices, in turn, trigger the Schwinger mechanism, and allow for the gauge-invariant generation of an effective gluon mass. In this particular approach, the gluon mass is directly related to quantities that are intrinsic to the bound-state formation itself, such as the ``transition  amplitude''  and  the corresponding  ``bound-state   wave-function''. Specifically, a set of powerful relations discussed in the text, allows one to determine the dynamical evolution of the gluon mass through a Bethe-Salpeter equation, which controls the dynamics of the relevant wave-function. In addition, it is possible to demonstrate that the massless bound-state formalism is equivalent to the standard approach based on Schwinger-Dyson equations, thus establishing a formal connection between two different nonperturbative formalisms.}

\FullConference{From quarks and gluons to hadronic matter: A bridge too far?\\
		 2-6 September, 2013\\
		 European Centre for Theoretical Studies in Nuclear Physics and Related 
Areas (ECT*), Villazzano, Trento (Italy)}

\begin{document}

\section{Introduction}

The dynamical generation of an effective (momentum-dependent) gluon mass in Yang-Mills theories~\cite{Cornwall:1981zr} has attracted considerable attention over the past few years, as it furnishes a convincing theoretical framework in which virtually all the recent lattice findings~\cite{Cucchieri:2007md,Bogolubsky:2007ud,Bowman:2007du,Oliveira:2009eh,Iritani:2009mp} can be easily accommodated.  This inherently nonperturbative phenomenon is usually studied within the formal machinery of the Schwinger-Dyson equations (SDEs), supplemented with a set of fundamental guiding principles, which enable the emergence of ``massive'' (propagator) solutions, while preserving intact the gauge invariance of the theory~\cite{Aguilar:2008xm,Aguilar:2011ux}.

The most crucial theoretical ingredient for achieving this result, \textit{without} interferring with the gauge invariance of the theory (encoded by the BRST symmetry), is the existence of a set of special vertices, to be generically denoted by $V$, called \textit{pole vertices}. These vertices contain massless, longitudinally coupled poles, and must be added to the usual (fully dressed) vertices of the theory; they are responsible for the underlying mass generation mechanism, as they trigger a non-Abelian realization of the Schwinger mechanism~\cite{Schwinger:1962tn}. In addition, the massless poles they contain act as composite, longitudinally coupled Nambu-Goldstone bosons: they maintain the gauge invariance of the theory (preserving the form of the Ward identities (WIs) and the Slavnov-Taylor identities (STIs) even when a gluon mass is dynamically generated), and they cancel out from $S$-matrix elements. 

Recent studies indicate that the QCD dynamics can indeed generate longitudinally coupled composite (bound-state) massless poles, which subsequently give rise to the required vertices $V$~\cite{Aguilar:2011xe}. Building on this, we describe our recent work~\cite{Ibanez:2012zk}, where it has been established a precise quantitative connection between the fundamental ingredients composing the aforementioned pole vertices and the gluon mass itself. Specifically, we dissect the pole vertices and scrutinize the field-theoretic properties of their constituents, within the context of the ``massless bound-state formalism'', first introduced in some early seminal contributions to this subject~\cite{Jackiw:1973tr,Jackiw:1973ha,Cornwall:1973ts,Eichten:1974et,Poggio:1974qs}, and further developed in~\cite{Aguilar:2011xe}. The final outcome of this analysis is an alternative description of the dynamical gluon mass in terms of quantities appearing naturally in the physics of bound states, such as the ``transition amplitude'' and the ``bound-state wave function'', which turns out to be completely equivalent to the standard approach based on SDEs.

\section{The SDE of the gluon propagator}

%%%%%%%%%%%%%%%%%%%%%%%%%%%%%%%%%%%%%%%%%%%%%%%%%%%%%%%%%%%%%%%%%%%%%%%%%%%%%%%%%%%%%%%%%%%
\begin{figure}[t!]
\center{\includegraphics[scale=0.35]{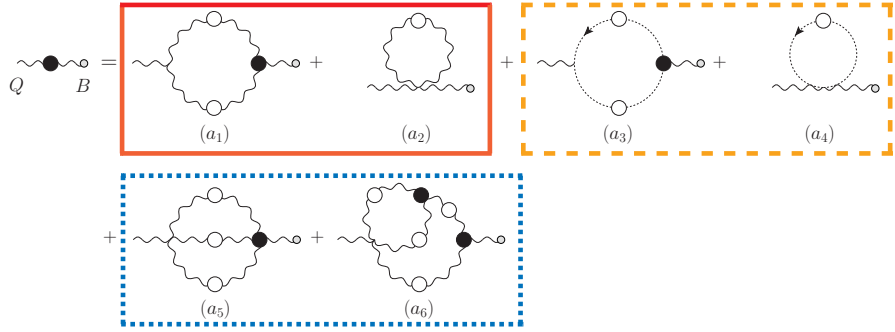}}
\caption{The SDE obeyed by the $QB$ gluon propagator. Black blobs represents fully dressed 1-PI vertices; the small gray circles appearing on the external legs are used to indicate background gluons.}
\label{QB-SDE}
\end{figure}
%%%%%%%%%%%%%%%%%%%%%%%%%%%%%%%%%%%%%%%%%%%%%%%%%%%%%%%%%%%%%%%%%%%%%%%%%%%%%%%%%%%%%%%%%%%
The full gluon propagator $\Delta^{ab}_{\mu\nu}(q)=\delta^{ab}\Delta_{\mu\nu}(q)$ in the Landau gauge is given by the expression
\begin{equation}\label{prop}
\Delta_{\mu\nu}(q)=-iP_{\mu\nu}(q)\Delta(q^2); \quad P_{\mu\nu}(q)=g_{\mu\nu}-\frac{q_\mu q_\nu}{q^2},
\end{equation}
and its \textit{inverse} gluon dressing function, $J(q^2)$, is defined as
\begin{equation}\label{gldressing}
\Delta^{-1}(q^2)=q^2 J(q^2).
\end{equation}
Within the framework provided by the synthesis of the pinch technique (PT) with the background field method (BFM), known in the literature as the PT-BFM scheme~\cite{Binosi:2002ft,Aguilar:2006gr}, one can consider the propagator connecting a quantum ($Q$) with a background ($B$) gluon, to be referred as the $QB$ propagator and denoted by $\widetilde{\Delta}(q^2)$. The SDE of the above propagator is shown in Fig.~\ref{QB-SDE}, and it may be related to the conventional $QQ$ propagator, $\Delta(q^2)$, connecting two quantum gluons, through the background-quantum identity~\cite{Binosi:2002ft,Grassi:1999tp}
\begin{equation}\label{propBQI}
\Delta(q^2) = [1+G(q^2)]\widetilde{\Delta}(q^2).
\end{equation}
In this identity, the function $G(q^2)$ corresponds to the $g_{\mu\nu}$ form factor of a special two-point function typical of the PT-BFM framework~\cite{Binosi:2002ft,Grassi:1999tp,Binosi:2007pi}; in the Landau gauge, it can be related to the dressing function $F(q^2)$ of the ghost propagator $D(q^2)=F(q^2)/q^2$ throughout the approximate relation
\begin{equation}\label{GF}
F^{-1}(q^2) \approx 1 + G(q^2),
\end{equation}
which becomes exact in the limit $q^2\rightarrow 0$~\cite{Aguilar:2009nf}. Then, the corresponding version of the SDE for the conventional gluon propagator (in the Landau gauge) reads~\cite{Aguilar:2006gr,Binosi:2007pi}
\begin{equation}\label{glSDE}
\Delta^{-1}(q^2)P_{\mu\nu}(q) = \frac{q^2P_{\mu\nu}(q) + i\sum_{i=1}^6(a_i)_{\mu\nu}}{1 + G(q^2)},
\end{equation}
where the diagrams $(a_i)$ are shown in Fig.~\ref{QB-SDE}. The relevant point to recognize here is that the transversality of the gluon self-energy is realized according to the pattern highlighted by the boxes of Fig.~\ref{QB-SDE}, namely,
\begin{equation}\label{pattern}
q^\mu[(a_1)+(a_2)]_{\mu\nu}=0; \quad q^\mu[(a_3)+(a_4)]_{\mu\nu}=0; \quad q^\mu[(a_5)+(a_6)]_{\mu\nu}=0.
\end{equation}

\section{SDE formalism}

As has been explained in detail in the recent literature~\cite{Aguilar:2011ux,Aguilar:2011xe}, the Schwinger mechanism allows for the emergence of massive solutions out of the SDE, preserving, at the same time, the gauge invariance intact. At this level, the triggering of this mechanism proceeds through the inclusion of the pole vertices $V$ in the SDE Eq.~(\ref{glSDE}). From the kinematic point of view, one can describe the transition from a massless to a massive gluon propagator by carrying out the replacement (Minkowski space)
\begin{equation}\label{massiveprop}
\Delta^{-1}(q^2)=q^2 J(q^2) \longrightarrow \Delta_m^{-1}(q^2) = q^2 J_m(q^2) - m^2(q^2).
\end{equation}
Notice that the subscript ``m'' indicates that effectively one has now a mass inside the corresponding expressions. Then, gauge invariance requires that the replacement given in Eq.~(\ref{massiveprop}) be accompanied by the following simultaneous replacement of all relevant vertices
\begin{equation}\label{replacever}
\Gamma \longrightarrow \Gamma' = \Gamma_m + V,
\end{equation}
where $V$ must be such that the new vertex $\Gamma'$ satisfies the same formal WIs (or STIs) as $\Gamma$ before. To see how this works with an explicit example consider the fully dressed vertex $BQ^2$ connecting a background gluon with two quantum gluons, to be denoted by $\widetilde{\Gamma}_{\alpha\mu\nu}$. With the Schwinger mechanism ``turned off'', this vertex satisfies the WI
\begin{equation}
q^\alpha\widetilde{\Gamma}_{\alpha\mu\nu}(q,r,p)=p^2J(p^2)P_{\mu\nu}(p)-r^2J(r^2)P_{\mu\nu}(r),
\label{STI}
\end{equation}
when contracted with respect to the momentum of the background gluon. The general replacement described in Eq.~(\ref{replacever}) amounts to introducing the vertex  
\begin{equation}
\widetilde{\Gamma}'_{\alpha\mu\nu}(q,r,p) = \left[\widetilde{\Gamma}_{m}(q,r,p) + \widetilde{V}(q,r,p)\right]_{\alpha\mu\nu};
\label{NV}
\end{equation}
then, gauge invariance requires that 
\begin{equation}
q^\alpha \widetilde{V}_{\alpha\mu\nu}(q,r,p)= m^2(r^2)P_{\mu\nu}(r) - m^2(p^2)P_{\mu\nu}(p),
\label{winp}
\end{equation}
so that, after turning the Schwinger mechanism on, the corresponding WI satisfied by $\widetilde{\Gamma}'$ would read  
\begin{eqnarray}
q^{\alpha}\widetilde{\Gamma}'_{\alpha\mu\nu}(q,r,p) &=& q^{\alpha}\left[\widetilde{\Gamma}_{m}(q,r,p) + \widetilde{V}(q,r,p)\right]_{\alpha\mu\nu} \nonumber \\
&=& [p^2 J_m (p^2) - m^2(p^2)]P_{\mu\nu}(p) - [r^2 J_m (r^2) - m^2(r^2)]P_{\mu\nu}(r) \nonumber\\
&=& \Delta_m^{-1}({p^2})P_{\mu\nu}(p) - \Delta_m^{-1}({r^2})P_{\mu\nu}(r),
\label{winpfull}
\end{eqnarray}
which is indeed the identity in Eq.~(\ref{STI}), with the aforementioned replacement \mbox{$\Delta^{-1} \to \Delta_m^{-1}$} enforced. In addition, the pole vertices must be completely \textit{longitudinally} coupled; i.e., they satisfy conditions of the type (for the case of the $BQ^2$ pole vertex)\footnote{Notice that this longitudinality condition implies that, in the Landau gauge, the pole vertices cannot be observed in lattice simulations of connected Green's functions.}
\begin{equation}\label{long}
P^{\alpha'\alpha}(q)P^{\mu'\mu}(r)P^{\nu'\nu}(p) \widetilde{V}_{\alpha'\mu'\nu'}(q,r,p) = 0.
\end{equation}

%%%%%%%%%%%%%%%%%%%%%%%%%%%%%%%%%%%%%%%%%%%%%%%%%%%%%%%%%%%%%%%%%%%%%%%%%%%%%%%%%%%%
\begin{figure}[t!]
\center{\includegraphics[scale=0.45]{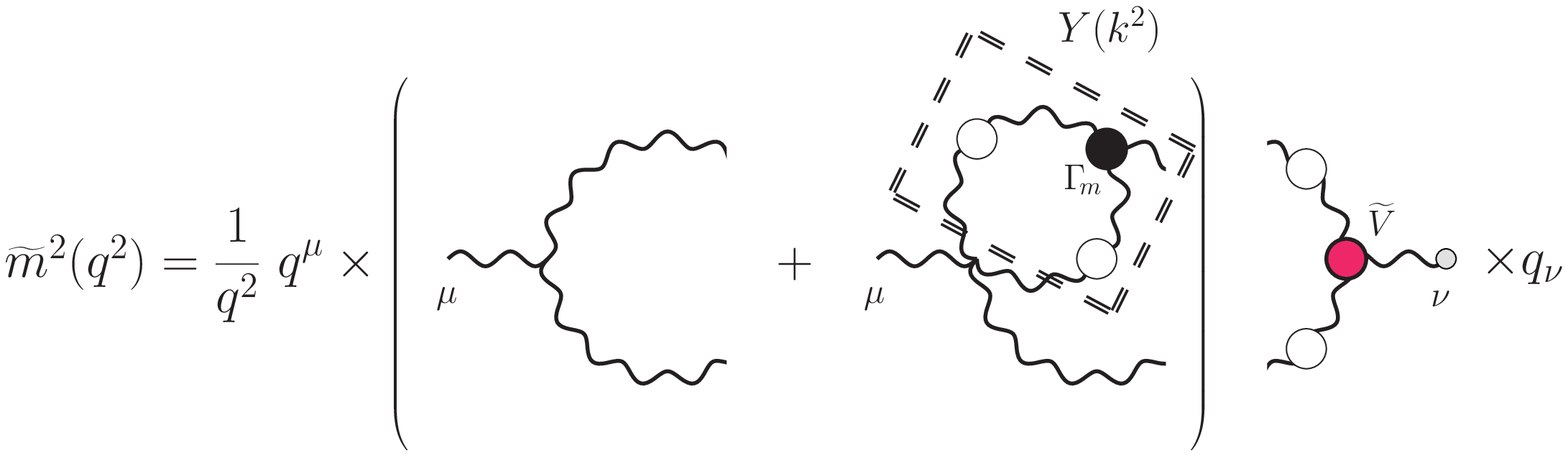}}
\caption{Diagrammatic representation of the condensed operations leading to the all-order gluon mass equation, where we have introduced the shorthand notation $\widetilde{m}^2(q^2)=m^2(q^2)[1+G(q^2)]$.}
\label{diagrammaticmass}
\end{figure}
%%%%%%%%%%%%%%%%%%%%%%%%%%%%%%%%%%%%%%%%%%%%%%%%%%%%%%%%%%%%%%%%%%%%%%%%%%%%%%%%%%%%

According to the previous discussion, after the inclusion of the pole vertices, the gluon SDE Eq.~(\ref{glSDE}) becomes in the Landau gauge
\begin{equation}\label{glSDEprime}
[q^2J_m(q^2) - m^2(q^2)]P_{\mu\nu}(q) = \frac{q^2P_{\mu\nu}(q) + i\sum_{i=1}^6(a'_i)_{\mu\nu}}{1 + G(q^2)},
\end{equation}
where the \textit{prime} indicates that (in general) one must perform the simultaneous replacements Eq.~(\ref{massiveprop}) and Eq.~(\ref{replacever}) inside the corresponding diagrams. At this point, following the detailed procedure presented in~\cite{Binosi:2012sj}, one is able to isolate the mass function in Eq.~(\ref{glSDEprime}) using as only input the WIs supplemented by the totally longitudinal nature of the pole vertices. The resulting mass equation reads (Minkowsky space)
\begin{equation}\label{masseq}
m^2(q^2) = \frac{ig^2C_A}{1+G(q^2)}\frac{1}{q^2}\int_k m^2(k^2) \Delta_{\gamma\rho}(k)\Delta_\mu^\rho(k+q){\cal K}_{SD}^{\gamma\mu}(q,k),
\end{equation}
where we have defined the SD kernel
\begin{eqnarray}\label{SDKernel}
{\cal K}_{SD}^{\gamma\mu}(q,k) &=& g^{\gamma\mu}[(k+q)^2 - k^2]\bigg\lbrace 1 + \frac{3}{4}ig^2C_A[Y(k+q)+Y(k)]\bigg\rbrace \nonumber \\
&+& \frac{3}{4}ig^2C_A(q^2g^{\gamma\mu} - 2q^\gamma q^\mu)[Y(k+q) - Y(k)],
\end{eqnarray}
which contains the purely two-loop dressed loop integral
\begin{equation}\label{Yintegral}
Y(k^2)=\frac{1}{d-1}\frac{k_\alpha}{k^2}\int_l \Delta^{\alpha\rho}(l)\Delta^{\beta\sigma}(l+k)\Gamma_{\sigma\rho\beta}.
\end{equation}
Interestingly enough, the entire procedure for deriving the mass equation may be pictorially summarized, in a rather concise way, as shown in Fig.~\ref{diagrammaticmass}. Then, under certain approximations, Eq.~(\ref{masseq}) can be solved numerically for the entire range of physical momenta, revealing the existence of positive-definite and monotonically decreasing gluon masses (see Fig.~\ref{solutions}).

\section{Massless bound state formalism}

Whereas in the SDE approach outlined in the previous section one relies predominantly on the global properties of the vertices $V$, within the massless bound-state formalism~\cite{Ibanez:2012zk} one takes, instead, a closer look at the field-theoretic composition of these vertices, establishing fundamental relations between their internal ingredients and the gluon mass itself. This becomes possible thanks to the key observation that, since the fully dressed vertices appearing in the diagrams of Fig.~\ref{QB-SDE} are themselves governed by their own SDEs, the appearance of such massless poles must be associated with very concrete modifications in the various structures composing them. 
%%%%%%%%%%%%%%%%%%%%%%%%%%%%%%%%%%%%%%%%%%%%%%%%%%%%%%%%%%%%%%%%%%%%%%%%%%%%%%%%%%%%
\begin{figure}[t!]
\center{\includegraphics[scale=0.45]{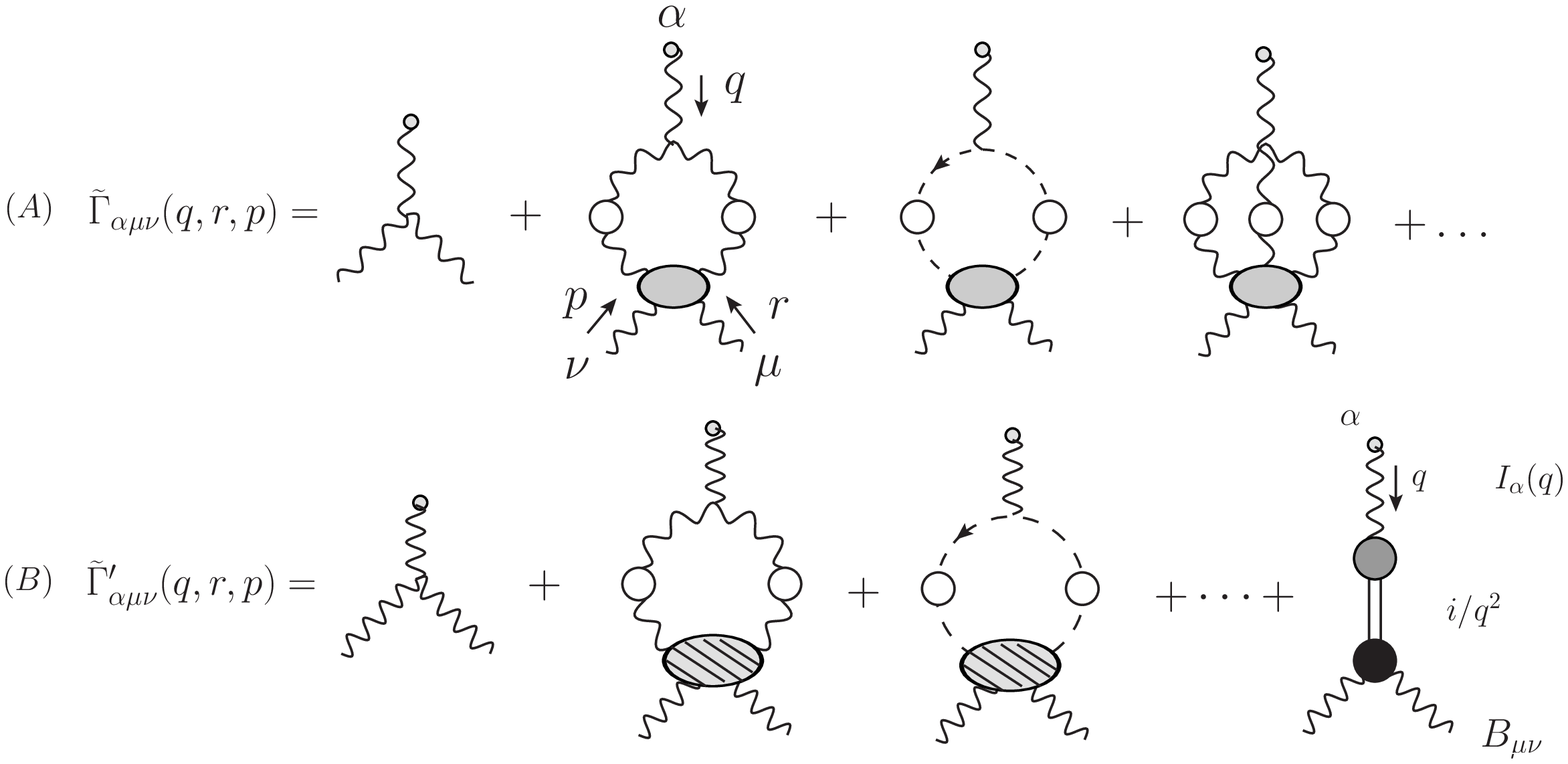}}
\caption{(A) The SDE for the $BQ^2$ vertex $\widetilde{\Gamma}_{\alpha\mu\nu}$. Gray blobs denote the conventional one-particle irreducible (with respect to vertical cuts) multiparticle kernels. (B) The SDE of the $BQ^2$ vertex in the presence of its pole part. The new SD kernels, which are modified with respect to those appearing in line (A), are described in Fig.~\protect\ref{Modifiedkernel}.}
\label{modifications}
\end{figure}
%%%%%%%%%%%%%%%%%%%%%%%%%%%%%%%%%%%%%%%%%%%%%%%%%%%%%%%%%%%%%%%%%%%%%%%%%%%%%%%%%%%%

\begin{figure}[t!]
\center{\includegraphics[scale=0.4]{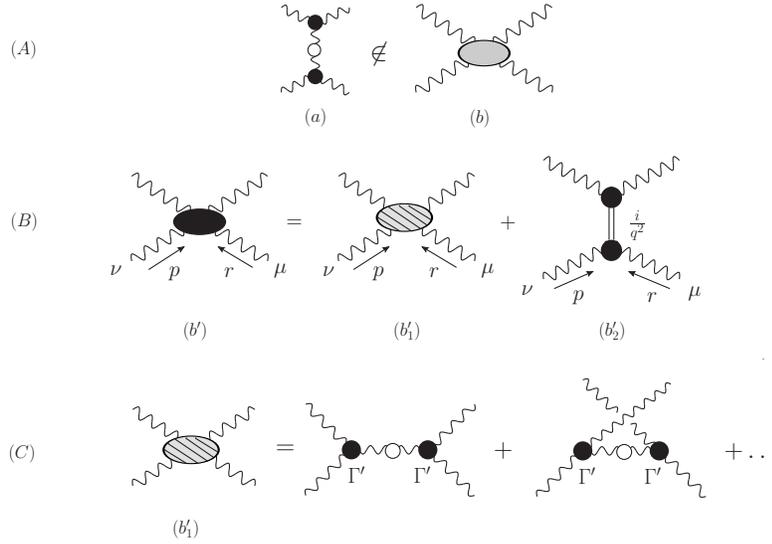}}
\caption{$(A)$ A diagram not included into the standard kernel. $(B)$ The kernel with the Schwinger mechanism turned on: in addition to the ``regular part'' $(b'_1)$, the massless excitation in the $q$-channel $(b'_2)$ is added. $(C)$ The part $(b'_1)$ is obtained from the original kernel $(b)$ by inserting massive gluon propagators into its diagrams, and carrying out the substitution Eq.~(\protect\ref{replacever}) in the fully-dressed vertices of the skeleton expansion.}
\label{Modifiedkernel}
\end{figure}

To fix the ideas we will focus our discussion on the case of the $BQ^2$ pole vertex $\widetilde{V}_{\alpha\mu\nu}$; however, all basic arguments may be straightforwardly extended to any generic pole vertex. Let us begin by recalling that, when the Schwinger mechanism is turned off, the various multiparticle kernels appearing in the SDE for the $BQ^2$ vertex, shown in line $(A)$ of Fig.~\ref{modifications}, have a complicated skeleton expansion, but their common characteristic is that they are {\it one-particle irreducible} with respect to cuts in the direction of the momentum $q$. Thus, for example, diagram $(a)$ of Fig.~\ref{Modifiedkernel} is explicitly excluded from the four-gluon kernel $(b)$, and the same is true for all other kernels. 

\begin{figure}[t!]
\center{\includegraphics[scale=0.5]{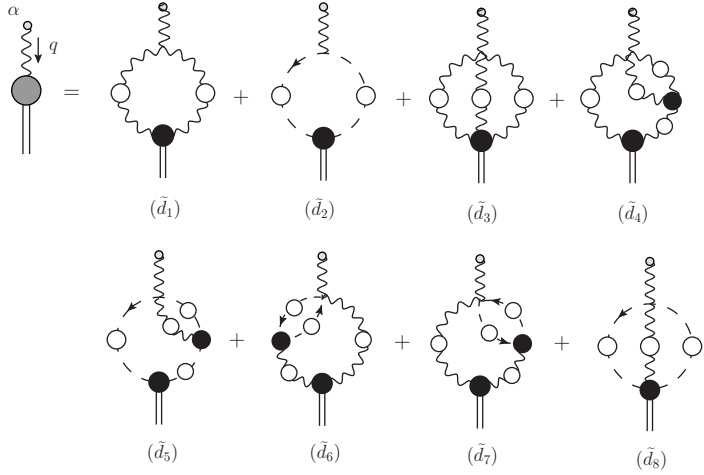}}
\caption{Diagrammatic representation of the background transition amplitude $\widetilde{I}_\alpha(q)$.}
\label{I}
\end{figure}

Now, when the Schwinger mechanism is turned on, the structure of the kernels is modified by the presence of the composite massless excitations, described by a propagator of the type $i/q^2$. For example, as shown in Fig.~\ref{Modifiedkernel}, the four-gluon kernel $(b)$ is converted into the kernel $(b')$, which is the sum of two parts: (i) the term $(b'_1)$, which corresponds to a kernel that is ``regular'' with respect to the $q$-channel, and (ii) the term $(b'_2)$, which describes the exchange of the composite massless excitation between two gluons in the $q$-channel. Thus, when the  replacements of Eq.~(\ref{massiveprop}) and Eq.~(\ref{replacever}) are carried out, the SDE for the $BQ^2$ vertex $\widetilde{\Gamma}'_{\alpha\mu\nu}$ in the presence of its pole part will be given by an expansion such as that shown in line $(B)$ of Fig.~\ref{modifications}.
%%%%%%%%%%%%%%%%%%%%%%%%%%%%%%%%%%%%%%%%%%%%%%%%%%%%%%%%%%%%%%%%%%%%%%%%%%%%%%%%%
%%%%%%%%%%%%%%%%%%%%%%%%%%%%%%%%%%%%%%%%%%%%%%%%%%%%%%%%%%%%%%%%%%%%%%%%%%%%%%%%%
These modifications in the composition of the kernels give rise precisely to the $BQ^2$ pole vertex $\widetilde{V}_{\alpha\mu\nu}$, which, in this formalism, is separated into two distinct parts, namely,
\begin{equation} \label{Vsep}
\widetilde{V}_{\alpha\mu\nu}(q,r,p) = \widetilde{{\cal U}}_{\alpha\mu\nu}(q,r,p) + \widetilde{{\cal R}}_{\alpha\mu\nu}(q,r,p),
\end{equation}
defined as follows. $\widetilde{{\cal U}}$ is the part of $\widetilde{V}$ that has its Lorentz index $\alpha$ saturated by the momentum $q$; thus, it necesarily contains the explicit $q$-channel massless excitation, namely, the $1/q^2$ poles. A closer look at the structure of the last diagram in line $(B)$ of Fig.~\ref{modifications} reveals that this is precisely the term to be identified with the $\widetilde{{\cal U}}$ part, which will be cast in the form
\begin{equation}\label{Upart}
\widetilde{{\cal U}}_{\alpha\mu\nu}(q,r,p) = \widetilde{I}_\alpha(q)\frac{i}{q^2}B_{\mu\nu}(q,r,p); \quad \widetilde{I}_\alpha(q)=q_\alpha \widetilde{I}(q^2).
\end{equation}
In this expression $\widetilde{I}_\alpha(q)$ denotes the background transition amplitude that mixes a background gluon with the massless excitation, whose diagrammatic expansion is shown in Fig.~\ref{I}. The quantity $i/q^2$ corresponds to the propagator of the massless excitation, and $B$ is an effective vertex describing the interaction between the massless excitation and gluons and/or ghosts. This latter vertex can be decomposed in a suitable tensor basis as follows
\begin{equation}\label{BLorentz}
B_{\mu\nu}(q,r,p) = B_1 g_{\mu\nu} + B_2 q_\mu q_\nu + B_3 p_\mu p_\nu + B_4 r_\mu q_\nu + B_5 r_\mu p_\nu,
\end{equation}
and, in the standard language used in bound-state physics, represents the bound-state wave function (or Bethe-Salpeter wave function), whose dynamics is controlled by a Bethe-Salpeter equation~\cite{Aguilar:2011xe}. Finally, the term $\widetilde{{\cal R}}$ contains everything else; in particular the massless excitations in the other two kinematic channels, namely, $1/r^2$ and $1/p^2$, are assigned to $\widetilde{{\cal R}}$.

Let us finally briefly discussing the two fundamental relations of the massless bound-state formalism, derived in~\cite{Ibanez:2012zk}. The first of them relates the gluon mass with the square of the transition amplitude and it emerges when the corresponding pole parts of the vertices appearing in the SDE of the gluon propagator are inserted in Eq.~(\ref{glSDEprime}). It simply reads
\begin{equation}\label{massrel}
m^2(q^2) = g^2 I^2(q^2),
\end{equation}
and demonstrates that, unless $I(q^2)$ vanishes identically, the gluon mass obtained is positive-definite. The second relation is obtained when one combines the WI Eq.~(\ref{winp}) with Eq.~(\ref{Upart}) and links the $g_{\mu\nu}$ form factor of the effective vertex $B_{\mu\nu}$ directly to the gluon mass through
\begin{equation}\label{IBmassrel}
[1+G(q^2)]I(q^2)B_1(q,r,p) = m^2(p^2) - m^2(r^2),
\end{equation}
which serves as the starting equation for obtaining the momentum dependence of the gluon mass in the formalism described here.
%%%%%%%%%%%%%%%%%%%%%%%%%%%%%%%%%%%%%%%%%%%%%%%%%%%%%%%%%%%%%%%%%%%%%%%%%%%%%%%%%%%%
%%%%%%%%%%%%%%%%%%%%%%%%%%%%%%%%%%%%%%%%%%%%%%%%%%%%%%%%%%%%%%%%%%%%%%%%%%%%%%%%%%%%
\section{Comparison of the two mass-generating formalisms}

At this point we have established two seemingly different formalisms for describing the phenomenon of gluon mass generation, namely, the SDE formalism and the massless bound-state formalism. Nonetheless, Eqs.~(\ref{massrel}) and~(\ref{IBmassrel}), can be used to demonstrate the exact coincidence of the integral equations governing the momentum evolution of the gluon mass in both frameworks. 

The demonstration begins with the evaluation of the transition amplitude which, in the Landau gauge, is given solely by the sum of diagrams $(\widetilde{d}_1)$ and $(\widetilde{d}_4)$ shown in Fig.~\ref{I}. Then, following the steps detailed in~\cite{Ibanez:2012zk}, one concludes that the \textit{complete} transition amplitude may be written as
\begin{eqnarray}\label{IB1general}
I(q^2) &=& \frac{i}{q^2}C_A\int_k k^2\Delta_\mu^\rho(k)\Delta^\mu_\rho(k+q)B_1 \nonumber \\
&+& \frac{3}{2}\frac{g^2C_A^2}{q^2}\int_k[(kq)g_{\mu\gamma} + q_\mu q_\gamma]Y(k^2)\Delta^\mu_\rho(k+q)\Delta^{\rho\gamma}(k)B_1,
\end{eqnarray}
which, quite remarkably, allows to express the full transition amplitude, for general value of $q^2$, in terms of one single form factor, namely, $B_1$. Now, we arrive at the crucial observation which enables us to relate the mass equation Eq.~(\ref{masseq}) obtained in the context of the PT-BFM formalism with the bound-state formalism showing that, indeed both formalisms are self-consistent and interconnected. Using Eq.~(\ref{IBmassrel}) and after a straightforward rearrangement, Eq.~(\ref{IB1general}) yields
\begin{equation}\label{squareI}
I^2(q^2) = \frac{iC_A}{1+G(q^2)}\frac{1}{q^2}\int_k \Delta_{\gamma\rho}(k)\Delta_\mu^\rho(k+q){\cal K}_{SD}^{\gamma\mu}(q,k)m^2(k^2),
\end{equation}
where we observe that the SD kernel defined in Eq.~(\ref{SDKernel}) appears. Thus, using the mass formula Eq.~(\ref{massrel}) on the lhs of Eq.~(\ref{squareI}), we reproduce {\it exactly} the full mass equation Eq.~(\ref{masseq}) derived in~\cite{Binosi:2012sj}, providing a striking self-consistency check between these two formalisms. However it is important to recognize that, although formally equivalent, the two approaches (``SDE'' vs ``massless bound-state'') entail vastly different procedures for obtaining the desired quantity, namely the functional form of $m^2 (q^2)$. Given that in practice approximations must be carried out to the fundamental equations of both formalisms, the results obtained for $m^2 (q^2)$ will be in general different; indeed, the formal coincidence proved above is only valid when all equations involved are treated exactly.

%%%%%%%%%%%%%%%%%%%%%%%%%%%%%%%%%%%%%%%%%%%%%%%%%%%%%%%%%%%%%%%%%
\begin{figure}[t!]
\center{\includegraphics[scale=0.48]{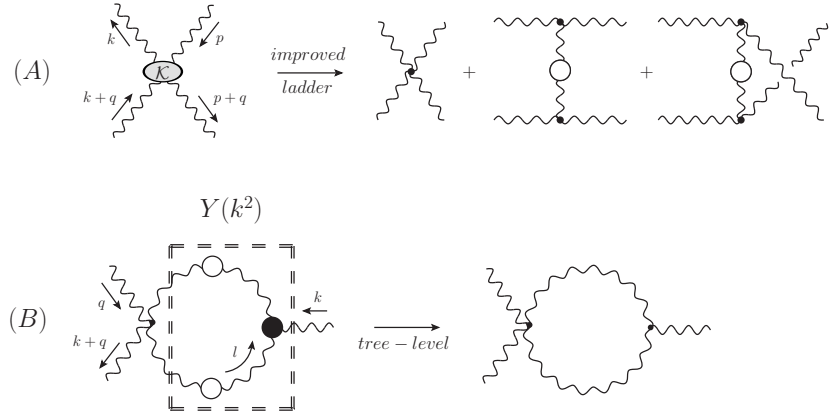}}
\caption{(A) The kernel ${\cal K}$ of the Bethe-Salpeter equation satisfied by $B_1'$ and the approximation introduced in~\cite{Aguilar:2011xe}. (B) The full kernel ${\cal K}_{SD}$ of the mass equation, and the approximation employed in~\cite{Binosi:2012sj}.}
\label{SDEvsBS}
\end{figure}
%%%%%%%%%%%%%%%%%%%%%%%%%%%%%%%%%%%%%%%%%%%%%%%%%%%%%%%%%%%%%%%%%

\begin{figure}[!t]
%\hspace{.1cm}
\begin{minipage}[b]{0.45\linewidth}
%\centering
\includegraphics[scale=0.4]{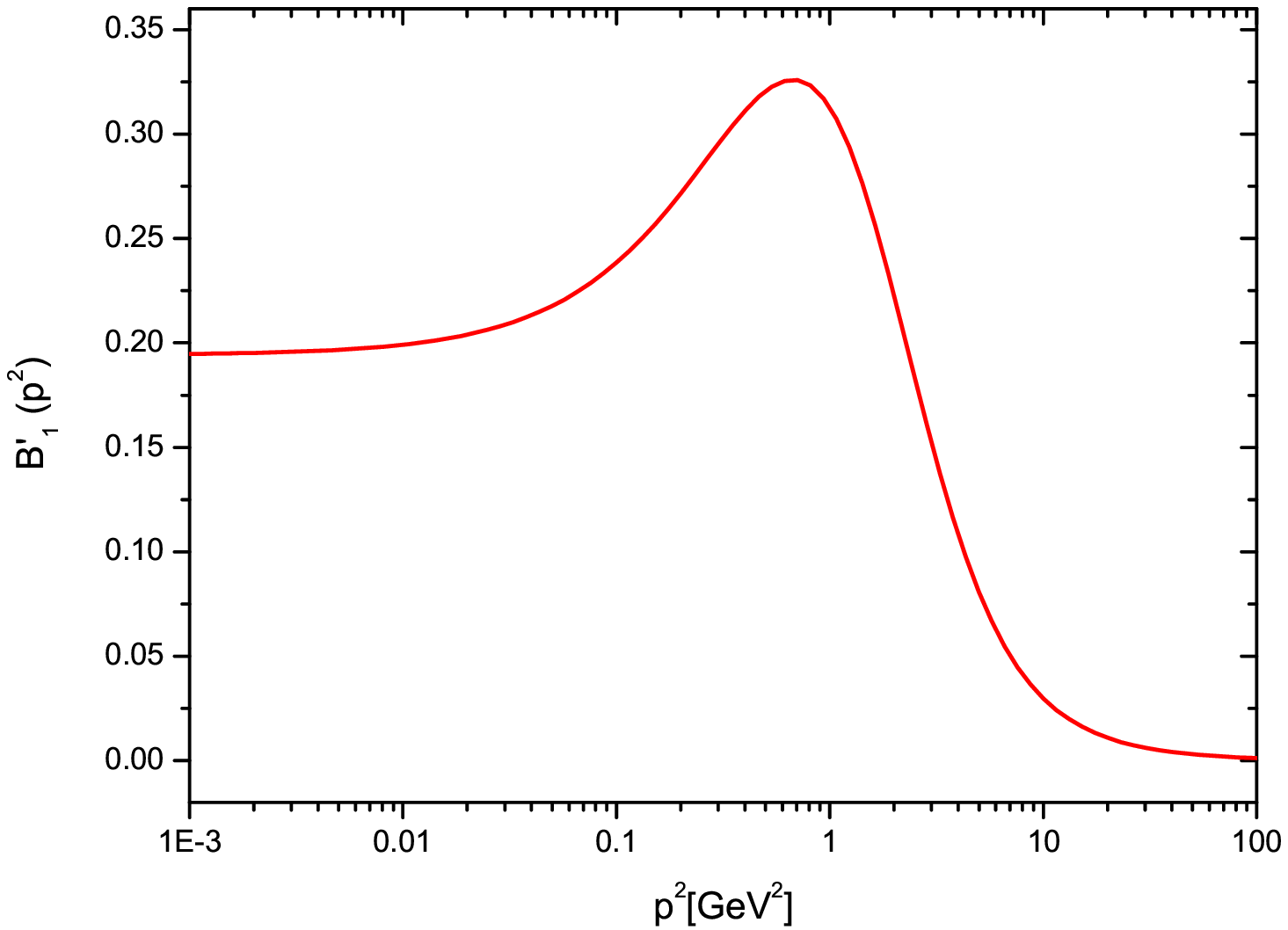}
\end{minipage}
\hspace{0.58cm}
\begin{minipage}[b]{0.50\linewidth}
\hspace{-1cm}
\includegraphics[scale=0.4]{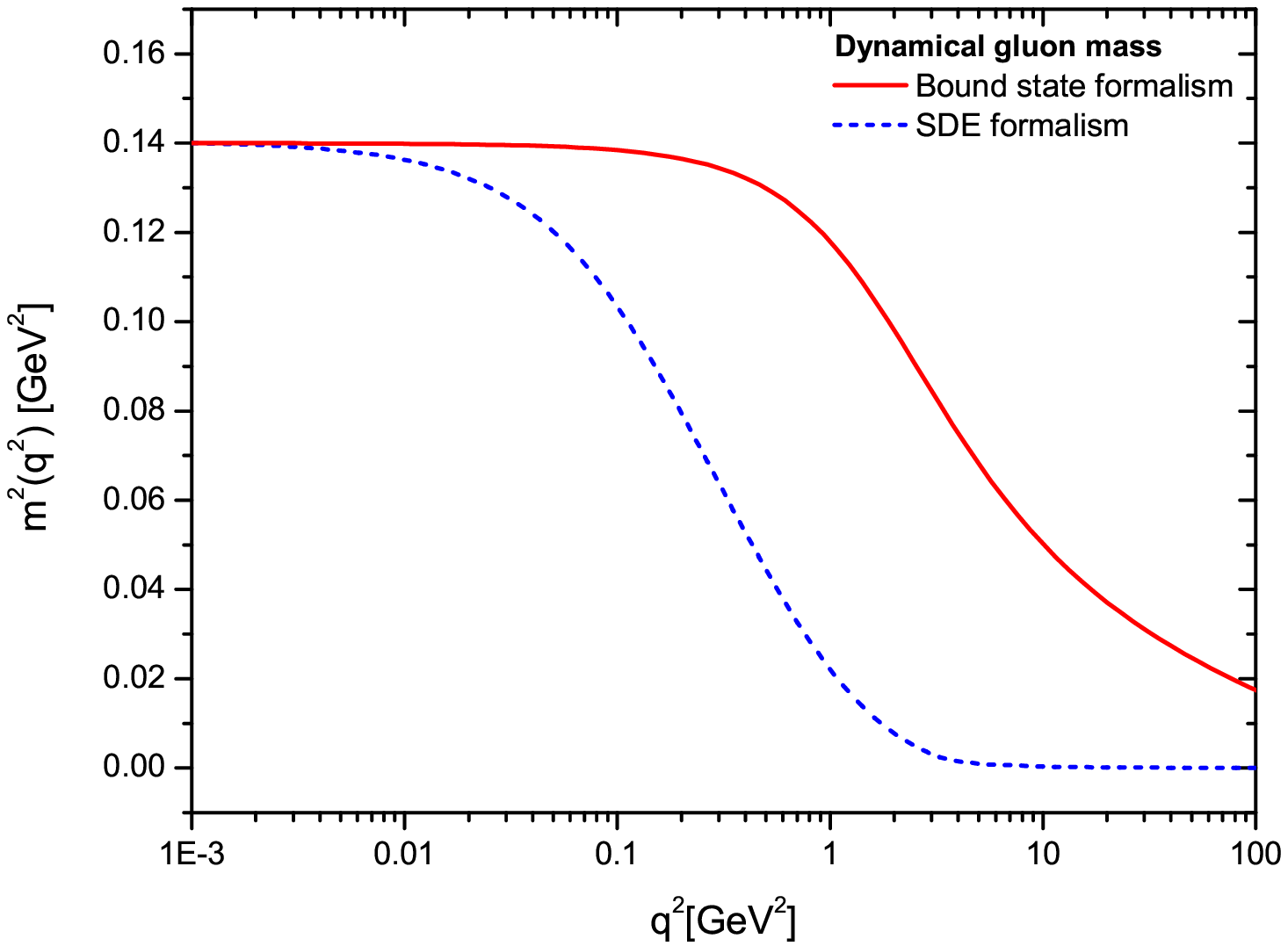}
\end{minipage}
\vspace{-0.25cm}
\caption{\label{solutions} (Left panel): 
The general form of $B'_1(p^2)$ obtained from the numerical solution of Eq.~(\protect\ref{BSEB1}), in the improved ladder approximation.
(Right panel): The two solutions obtained for the gluon mass within the massless bound-state [red (solid)] and SDE [blue (dotted)] approaches}
\end{figure}

To appreciate this issue in some quantitative detail, recall that, within the massless bound-state formalism, the momentum-dependence of the gluon mass is obtained from Eq.~(\ref{IBmassrel}). Specifically, in the limit $q\rightarrow 0$ one obtains the 
following exact relation for the derivative of the effective gluon mass (Euclidean space)~\cite{Aguilar:2011xe}
\begin{equation}\label{massderivative}
\frac{\mathrm{d} m^2(p^2)}{\mathrm{d} p^2} = -F^{-1}(0)I(0)B_1'(p^2); \quad B_1'(p^2) \equiv \lim_{q\to 0} \left\{ \frac{\partial B_1(q,-p-q,p)}{\partial (p+q)^2} \right\}.
\end{equation}
It turns out that the function $B_1'(p^2)$ satisfies its own homogeneous Bethe-Salpeter equation, of the general form~\cite{Aguilar:2011xe}
\begin{equation}\label{BSEB1}
B_1'(p^2) = \int_k {\cal K}(p,k) B_1'(k^2), 
\end{equation}
where ${\cal K}$ corresponds to the Bethe-Salpeter four-gluon kernel, shown in line $(A)$ of Fig.~\ref{SDEvsBS}. Evidently, the exact treatment of this equation would require the complete knowledge of the four-gluon kernel, which is a largely unexplored quantity. Therefore, in order to obtain an approximate solution to Eq.~(\ref{BSEB1}), one resorts to the ``improved'' ladder approximation of this kernel, shown diagrammatically in Fig.~\ref{SDEvsBS}, by dressing the gluon propagators but keeping the vertices at tree-level. Then, the numerical solution of Eq.~(\ref{BSEB1}) yields for $B_1'(q^2)$ the function shown on the left panel of Fig.~\ref{solutions}. With this information at hand, the  gluon mass may be obtained through direct integration of Eq.~(\ref{massderivative}), namely 
\begin{equation}\label{massderB1}
m^2(q^2) = m^2(0) - F^{-1}(0)I(0) \int_0^{q^2} dx B_1'(x).
\end{equation}
As for the constant $F^{-1}(0)I(0)$, using Eq.~(\ref{massrel}), one obtains 
\begin{equation}\label{valuezerotran}
F^{-1}(0)I(0) = F^{-1}(0)\sqrt{\frac{m^2(0)}{g^2}} = F^{-1}(0)\sqrt{\frac{\Delta^{-1}(0)}{4\pi\alpha_s}},
\end{equation}
which allow us to estimate $I(0)$ from the lattice values of the ghost dressing function and the gluon propagator at zero momentum, treating $\alpha_s$ as an adjustable parameter. Thus, one finally arrives at the dynamical gluon mass shown on the right panel of Fig.~\ref{solutions} (red continuous curve)~\cite{Aguilar:2011xe}.

In the case of the SDE approach, the basic dynamical equation is that of Eq.~(\ref{masseq}), whose central ingredient is the quantity $Y(k^2)$, entering into the kernel ${\cal K}_{SD}$, given in Eq.~(\ref{SDKernel}). The quantity $Y(k^2)$, shown in line $B$ of Fig.~\ref{SDEvsBS}, involves the {\it fully-dressed} three-gluon vertex, which too is rather poorly known. As a result, in~\cite{Binosi:2012sj}, $Y(k^2)$ has been approximated by its perturbative (one-loop) expression, by replacing the fully dressed internal gluon  propagators and three-gluon vertex by their tree-level values, as shown in  Fig.~\ref{SDEvsBS}. Under these approximations, the numerical treatment of Eq.~(\ref{masseq}) gives rise to the solution shown on the right panel of Fig.~\ref{solutions} (blue dotted curve).  

It is clear from this direct comparison that the two formally equivalent approaches lead to qualitatively similar results, which, however, do not coincide, due to the inequivalence of the approximations employed.

%%%%%%%%%%%%%%%%%%%%%%%%%%%%%%%%%%%%%%%%%%%%%%%%%%%%%%%%%%%%%%%%%
%%%%%%%%%%%%%%%%%%%%%%%%%%%%%%%%%%%%%%%%%%%%%%%%%%%%%%%%%%%%%%%%%

\section{Final remarks: About the decoupling of the massless poles}

Before presenting our conclusions, we add in this section some final remarks about the general mechanism that leads to the decoupling of all massless poles from the physical (on-shell) amplitudes of the theory. It is precisely this mechanism which explains how these Goldstone-like particles cancel in the $S$-matrix elements and why they cannot be measured in physical processes, neither be observed in the spectrum of the theory. Even though these issues have been addressed in detail in the past and recent literature~\cite{Jackiw:1973tr,Jackiw:1973ha,Cornwall:1973ts,Eichten:1974et,Poggio:1974qs}, they have come up in the discussion session of this presentation; as a result we think it is important to go briefly over them and clarify them as much as possible.

\begin{figure}[t!]
\center{\includegraphics[scale=0.4]{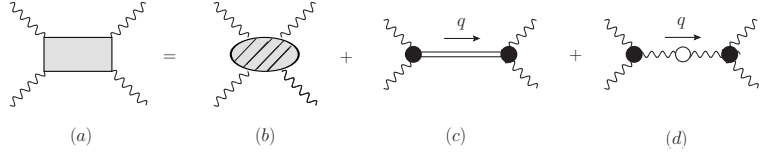}}
\caption{The decomposition of the complete four-gluon scattering amplitude.}
\label{decoupling}
\end{figure}

To fix the ideas, let us consider a specific process, and namely the one of the four-gluon scattering amplitude. As it is shown in Fig.~\ref{decoupling}, the \textit{complete} four-gluon scattering amplitude [graph $(a)$] is given by the sum of three different contributions, namely: (i) the amplitude $(b)$, which is regular as $q^2\to 0$, (ii) the graph $(c)$, which contains the massless excitation, coupled to the external gluons through the proper vertex function $B_{\mu\nu}$, and (iii) the one-particle reducible term, denoted by $(d)$, which is excluded from the SDE kernel in the usual skeleton expansion. Note that the above amplitudes are none other than $(b_1')$,  $(b_2')$, and $(a)$ in Fig.~\ref{Modifiedkernel}, respectively. Thus, since the amplitude $(b)$ is regular by construction, one must only demonstrate that, as $q^2\to 0$, the divergent part of $(c)$, whose origin is the massless excitation, cancels exactly against an analogous contribution contained in  $(d)$, leaving finally a regular result.
%%%%%%%%%%%%%%%%%%%%%%%%%%%%%%%%%%%%%%%%%%%%%%%%%%%%%%%%%%%%%%%%%
%%%%%%%%%%%%%%%%%%%%%%%%%%%%%%%%%%%%%%%%%%%%%%%%%%%%%%%%%%%%%%%%%
Effectively, the term $(d)$ with the replacements of Eqs.~(\ref{massiveprop}) and (\ref{replacever}) already implemented reads
\begin{eqnarray}\label{dec1}
(d) &=& -ig^2 \widetilde{\Gamma}'_{\alpha\mu\nu}(q,p_1,p_2)\widehat{\Delta}(q^2)P^{\alpha\beta}(q)\widetilde{\Gamma}'_{\beta\rho\sigma}(q,p_3,p_4) \nonumber \\
&=& -ig^2 \widetilde{\Gamma}'_{\alpha\mu\nu}(q,p_1,p_2)\widehat{\Delta}(q^2)\widetilde{\Gamma}^{'\alpha}_{\rho\sigma}(q,p_3,p_4),
\end{eqnarray}
where, within the PT-BFM framework that we use, the off-shell gluon (carrying momentum $q$) is converted into a background gluon, while the two three-gluon vertices are the $\widetilde{\Gamma}'$ defined in Eq.~(\ref{NV}). Note that in the second line we have eliminated the longitudinal term $q^\alpha q^\beta /q^2$ inside $P^{\alpha\beta}(q)$ using the ``on-shellness'' condition
\begin{equation}\label{onshell}
q^\alpha \widetilde{\Gamma}'_{\alpha\mu\nu}(q,p_1,p_2)\vert_{\rm o.s.} = 0,
\end{equation}
valid for both three-gluon vertices. Then, following the demonstration outlined in~\cite{Aguilar:2011xe}, one may isolate the divergent contribution of the massless pole $1/q^2$ to Eq.~(\ref{dec1}), obtaining the result
\begin{equation}\label{dec2}
\lim_{q^2\to 0} (d)_{\rm pole} = - \lim_{q^2\to 0} \left\{B \left(\frac{i}{q^2}\right) B \right\},
\end{equation}
which is precisely the contribution of the term $(c)$ in the same kinematic limit, but with the opposite sign. Therefore, the on-shell four-gluon amplitude is free from poles at $q^2 = 0$, as announced. 

Finally, a general and process independent proof of this decoupling mechanism, involving an arbitrary on-shell scattering amplitude, may be found in~\cite{Eichten:1974et}. There, the generalization proceeds through the following two basic observations: (i) since in an arbitrary amplitude all the particles except the internal gluon (with momentum $q$) are on the mass shell, similar conditions to Eq.~(\ref{onshell}) remain valid, and (ii) the massless pole contributions can be factorized from the regular parts of the amplitude which, in the previous example, occurs when the replacement of Eq.~(\ref{replacever}) is carried out in diagram $(d)$.

\section{Conclusions}

In this presentation we have reported recent progress~\cite{Ibanez:2012zk} on the study of gluon mass generation within the massless bound-state formalism, which constitutes the formal framework for the systematic implementation of the Schwinger mechanism at the level of non-Abelian gauge theories. The main ingredient of this formalism is the dynamical formation of massless bound-states, which give rise to effective vertices containing massless poles; these latter vertices trigger the Schwinger mechanism, and allow for the gauge-invariant generation of an effective gluon mass. The principal advantage of this approach is its ability to relate the gluon mass directly to quantities that are usually employed in the physics of bound-states, such as transition amplitudes and bound-state wave functions, as well as obtaining the dynamical evolution through a Bethe-Salpeter equation instead of a SDE [see, Eqs.~(\ref{massrel}), (\ref{BSEB1}) and (\ref{massderB1})]. A central result is the formal equivalence between the massless bound-state formalism and the corresponding approach based on the direct study of the SDE of the gluon propagator~\cite{Binosi:2012sj}. In particular, the relations given in Eqs.~(\ref{massrel}) and (\ref{IBmassrel}), allow us to demonstrate the exact coincidence of the integral equations governing the momentum evolution of the gluon mass in both formalisms.

\textit{Acknowledgements:} I would like to thank the ECT* for their hospitality and for making this workshop possible.


\begin{thebibliography}{99}

%\cite{Cornwall:1981zr}
\bibitem{Cornwall:1981zr}
J.~M.~Cornwall,
%``Dynamical Mass Generation In Continuum QCD,''
Phys.\ Rev.\ D {\bf 26}, 1453 (1982).
%%CITATION = PHRVA,D26,1453;%%


%\cite{Cucchieri:2007md}
\bibitem{Cucchieri:2007md}
A.~Cucchieri and T.~Mendes,
%``What's up with IR gluon and ghost propagators in Landau gauge? A puzzling
%answer from huge lattices,''
PoS {\bf LAT2007}, 297 (2007);
%  [arXiv:0710.0412 [hep-lat]].
%%CITATION = POSCI,LAT2007,297;%%
%\cite{Cucchieri:2007rg}
%\bibitem{Cucchieri:2007rg}
%A.~Cucchieri and T.~Mendes,
%``Constraints on the IR behavior of the gluon propagator in Yang-Mills
%theories,''
Phys.\ Rev.\ Lett.\  {\bf 100}, 241601 (2008);
%  [arXiv:0712.3517 [hep-lat]].
%%CITATION = PRLTA,100,241601;%%  
%\cite{Cucchieri:2009zt}
%\bibitem{Cucchieri:2009zt}
%A.~Cucchieri and T.~Mendes,
%``Landau-gauge propagators in Yang-Mills theories at beta = 0: massive
%solution versus conformal scaling,''
Phys.\ Rev.\  D {\bf 81}, 016005 (2010);
%[arXiv:0904.4033 [hep-lat]].
%%CITATION = PHRVA,D81,016005;%%
%\cite{Cucchieri:2011ga}
%\bibitem{Cucchieri:2011ga}
%A.~Cucchieri and T.~Mendes,
%``Further Investigation of Massive Landau-Gauge Propagators in the Infrared
%Limit,''
PoS {\bf LATTICE2010}, 280 (2010);
%  [arXiv:1101.4537 [hep-lat]].
%%CITATION = POSCI,LATTICE2010,280;%%
%\cite{Cucchieri:2011um}
%\bibitem{Cucchieri:2011um}
% A.~Cucchieri and T.~Mendes,
 %``The Saga of Landau-Gauge Propagators: Gathering New Ammo,''
 AIP Conf.\ Proc.\  {\bf 1343}, 185 (2011).
% [arXiv:1101.4779 [hep-lat]].
 %%CITATION = APCPC,1343,185;%%

%\cite{Bogolubsky:2007ud}
\bibitem{Bogolubsky:2007ud}
I.~L.~Bogolubsky, E.~M.~Ilgenfritz, M.~Muller-Preussker and A.~Sternbeck,
%``The Landau gauge gluon and ghost propagators in 4D SU(3) gluodynamics in
%large lattice volumes,''
PoS {\bf LAT2007}, 290 (2007);
%  [arXiv:0710.1968 [hep-lat]].
%%CITATION = POSCI,LATTICE,290;%%
%\cite{Bogolubsky:2009dc}
%\bibitem{Bogolubsky:2009dc}
%I.~L.~Bogolubsky, E.~M.~Ilgenfritz, M.~Muller-Preussker and A.~Sternbeck,
%``Lattice gluodynamics computation of Landau gauge Green's functions in the
%deep infrared,''
Phys.\ Lett.\  B {\bf 676}, 69 (2009).
% [arXiv:0901.0736 [hep-lat]].
%%CITATION = PHLTA,B676,69;%%


%\cite{Bowman:2007du}
\bibitem{Bowman:2007du}
P.~O.~Bowman {\it et al.},
%``Scaling behavior and positivity violation of the gluon propagator 
%in full QCD,''
Phys.\ Rev.\  D {\bf 76}, 094505 (2007).
%  [arXiv:hep-lat/0703022].
%%CITATION = PHRVA,D76,094505;%%


%\cite{Oliveira:2009eh}
\bibitem{Oliveira:2009eh}
O.~Oliveira and P.~J.~Silva,
%``The lattice infrared Landau gauge gluon propagator: the infinite volume
%limit,''
PoS {\bf LAT2009}, 226 (2009).
% [arXiv:0910.2897 [hep-lat]].
%%CITATION = POSCI,LAT2009,226;%%


%\cite{Iritani:2009mp}
\bibitem{Iritani:2009mp} 
  T.~Iritani, H.~Suganuma and H.~Iida,
  %``Gluon-propagator functional form in the Landau gauge in SU(3) lattice QCD: Yukawa-type gluon propagator and anomalous gluon spectral function,''
  Phys.\ Rev.\ D {\bf 80}, 114505 (2009).
%  [arXiv:0908.1311 [hep-lat]].
  %%CITATION = ARXIV:0908.1311;%%


%\cite{Aguilar:2008xm}
\bibitem{Aguilar:2008xm} 
  A.~C.~Aguilar, D.~Binosi and J.~Papavassiliou,
  %``Gluon and ghost propagators in the Landau gauge: Deriving lattice results from Schwinger-Dyson equations,''
  Phys.\ Rev.\ D {\bf 78}, 025010 (2008).
  %[arXiv:0802.1870 [hep-ph]].
  %%CITATION = ARXIV:0802.1870;%%


%\cite{Aguilar:2011ux}
\bibitem{Aguilar:2011ux} 
  A.~C.~Aguilar, D.~Binosi and J.~Papavassiliou,
  %``The dynamical equation of the effective gluon mass,''
  Phys.\ Rev.\ D {\bf 84}, 085026 (2011).
  %[arXiv:1107.3968 [hep-ph]].
  %%CITATION = ARXIV:1107.3968;%%


%\cite{Schwinger:1962tn}
\bibitem{Schwinger:1962tn}
  J.~S.~Schwinger,
  %``GAUGE INVARIANCE AND MASS,''
  Phys.\ Rev.\  {\bf 125}, 397 (1962);
  %%CITATION = PHRVA,125,397;%%
%\cite{Schwinger:1962tp}
%\bibitem{Schwinger:1962tp}
  %J.~S.~Schwinger,
  %``Gauge Invariance And Mass. 2,''
  Phys.\ Rev.\  {\bf 128}, 2425 (1962).
  %%CITATION = PHRVA,128,2425;%%
  
  
%\cite{Aguilar:2011xe}
\bibitem{Aguilar:2011xe}
  A.~C.~Aguilar, D.~Ibanez, V.~Mathieu and J.~Papavassiliou,
  %``Massless bound-state excitations and the Schwinger mechanism in QCD,''
  Phys.\ Rev.\ D {\bf 85} (2012) 014018.
  %[arXiv:1110.2633 [hep-ph]].
  %%CITATION = ARXIV:1110.2633;%%


%\cite{Ibanez:2012zk}
\bibitem{Ibanez:2012zk}
  D.~Iba\~nez and J.~Papavassiliou,
  %``Gluon mass generation in the massless bound-state formalism,''
  Phys.\ Rev.\ D {\bf 87} (2013) 3,  034008
  %[arXiv:1211.5314 [hep-ph]].
  %%CITATION = ARXIV:1211.5314;%%
  %6 citations counted in INSPIRE as of 15 Jan 2014


%\cite{Jackiw:1973tr}
\bibitem{Jackiw:1973tr}
  R.~Jackiw and K.~Johnson,
  %``Dynamical Model Of Spontaneously Broken Gauge Symmetries,''
  Phys.\ Rev.\ D {\bf 8}, 2386 (1973).
  %%CITATION = PHRVA,D8,2386;%%


%\cite{Jackiw:1973ha}
\bibitem{Jackiw:1973ha}
  R.~Jackiw,
  %``Dynamical Symmetry Breaking,''
  In *Erice 1973, Proceedings, Laws Of Hadronic Matter*.
  

%\cite{Cornwall:1973ts}
\bibitem{Cornwall:1973ts}
  J.~M.~Cornwall and R.~E.~Norton,
  %``Spontaneous Symmetry Breaking Without Scalar Mesons,''
  Phys.\ Rev.\ D {\bf 8} 3338 (1973).
  %%CITATION = PHRVA,D8,3338;%%


%\cite{Eichten:1974et}
\bibitem{Eichten:1974et}
E.~Eichten and F.~Feinberg,
%``Dynamical Symmetry Breaking Of Nonabelian Gauge Symmetries,''
Phys.\ Rev.\ D {\bf 10}, 3254 (1974).
  %%CITATION = PHRVA,D10,3254;%%


%\cite{Poggio:1974qs}
\bibitem{Poggio:1974qs}
  E.~C.~Poggio, E.~Tomboulis and S.~H.~Tye,
  %``Dynamical Symmetry Breaking In Nonabelian Field Theories,''
  Phys.\ Rev.\  D {\bf 11}, 2839 (1975).
  %%CITATION = PHRVA,D11,2839;%%
  

%\cite{Binosi:2002ft}
\bibitem{Binosi:2002ft}
D.~Binosi and J.~Papavassiliou,
%``The pinch technique to all orders,''
Phys.\ Rev.\  D {\bf 66}(R), 111901 (2002);
%[arXiv:hep-ph/0208189].
%%CITATION = PHRVA,D66,111901;%%
%\cite{Binosi:2003rr}
%\bibitem{Binosi:2003rr}
%D.~Binosi and J.~Papavassiliou,
%``Pinch technique self-energies and vertices to all orders in perturbation
%theory,''
J.\ Phys.\ G {\bf 30}, 203 (2004);
%[arXiv:hep-ph/0301096].
%%CITATION = JPHGB,G30,203;%%
%\cite{Binosi:2009qm}
%\bibitem{Binosi:2009qm} 
%  D.~Binosi and J.~Papavassiliou,
  %``Pinch Technique: Theory and Applications,''
  Phys.\ Rept.\  {\bf 479}, 1 (2009).
%  [arXiv:0909.2536 [hep-ph]].
  %%CITATION = ARXIV:0909.2536;%%


%\cite{Aguilar:2006gr}
\bibitem{Aguilar:2006gr}
  A.~C.~Aguilar and J.~Papavassiliou,
  %``Gluon mass generation in the PT-BFM scheme,''
  JHEP {\bf 0612} (2006) 012.
  %[hep-ph/0610040].
  %%CITATION = HEP-PH/0610040;%%


%\cite{Grassi:1999tp}
\bibitem{Grassi:1999tp}
  P.~A.~Grassi, T.~Hurth and M.~Steinhauser,
  %``Practical algebraic renormalization,''
  Annals Phys.\  {\bf 288} (2001) 197.
  %[hep-ph/9907426].
  %%CITATION = HEP-PH/9907426;%%


%\cite{Binosi:2007pi}
\bibitem{Binosi:2007pi}
  D.~Binosi and J.~Papavassiliou,
  %``Gauge-invariant truncation scheme for the Schwinger-Dyson equations of QCD,''
  Phys.\ Rev.\ D {\bf 77} (2008) 061702;
  %[arXiv:0712.2707 [hep-ph]].
  %%CITATION = ARXIV:0712.2707;%%
%\cite{Binosi:2008qk}
%\bibitem{Binosi:2008qk}
  %D.~Binosi and J.~Papavassiliou,
  %``New Schwinger-Dyson equations for non-Abelian gauge theories,''
  JHEP {\bf 0811} (2008) 063.
  %[arXiv:0805.3994 [hep-ph]].
  %%CITATION = ARXIV:0805.3994;%%
  

%\cite{Aguilar:2009nf}
\bibitem{Aguilar:2009nf}
  A.~C.~Aguilar, D.~Binosi, J.~Papavassiliou and J.~Rodriguez-Quintero,
  %``Non-perturbative comparison of QCD effective charges,''
  Phys.\ Rev.\ D {\bf 80} (2009) 085018.
  %[arXiv:0906.2633 [hep-ph]].
  %%CITATION = ARXIV:0906.2633;%%
  %95 citations counted in INSPIRE as of 21 Jan 2014
  

%\cite{Binosi:2012sj}
\bibitem{Binosi:2012sj}
  D.~Binosi, D.~Ibanez and J.~Papavassiliou,
  %``The all-order equation of the effective gluon mass,''
  Phys.\ Rev.\ D {\bf 86} (2012) 085033.
  %[arXiv:1208.1451 [hep-ph]].
  %%CITATION = ARXIV:1208.1451;%%
  %10 citations counted in INSPIRE as of 15 Jan 2014




\end{thebibliography}
\end{document}